# Excited-state optically detected magnetic resonance of spin defects in hexagonal boron nitride

Zhao Mu,[1,∆] Hongbing Cai,[1,∆] Disheng Chen,[1] Jonathan Kenny,[1] Zhengzhi Jiang,[2] Shihao Ru,[1] Xiaodan Lyu,[1] Teck Seng Koh,[1] Xiaogang Liu,[2] Igor Aharonovich,[3,4,‡] and Weibo Gao[1,5,†]

[1]*Division of Physics and Applied Physics, School of Physical and Mathematical Sciences, Nanyang Technological University, 637371, Singapore*
[2]*Department of Chemistry, National University of Singapore, 117543, Singapore*
[3]*School of Mathematical and Physical Sciences, University of Technology Sydney, Ultimo, New South Wales 2007, Australia*
[4]*ARC Centre of Excellence for Transformative Meta-Optical Systems, University of Technology Sydney, Ultimo, New South Wales 2007, Australia*
[5]*The Photonics Institute and Centre for Disruptive Photonic Technologies, Nanyang Technological University, 637371, Singapore*

*Negatively charged boron vacancy ($V_B^-$) centers in hexagonal boron nitride (hBN) are promising spin defects in a van der Waals crystal. Understanding the spin properties of the excited state (ES) is critical for realizing dynamic nuclear polarization (DNP). Here, we report zero-field splitting in the ES of $D_{ES}$ = 2160 MHz and its associated optically detected magnetic resonance (ODMR) contrast of 12% at cryogenic temperature. In contrast to nitrogen vacancy ($NV^-$) centers in diamond, the ODMR contrast is more prominent at cryo-temperature than at room temperature. The ES has a g-factor similar to the ground state. The ES photodynamics is further elucidated by measuring the level anti-crossing of the $V_B^-$ defects under varying external magnetic fields. Our results provide important information for utilizing the spin defects of hBN in quantum technology.*

Color centers with optically addressable spins in wide bandgap materials (e.g., diamond [1,2] and silicon carbide [3,4]) have been intensively studied in recent decades for applications in quantum sensing [5-8] and quantum information processing [9,10]. For nanoscale sensing, it is preferable to bring the sensor close to the investigated object to enhance sensitivity [11-13]. Spin qubits in 2D materials naturally meet this requirement and present an extra opportunity for quantum sensing besides the remarkable spatial resolution and sensitivity achieved by diamond $NV^-$ color centers [14,15].

Among various 2D materials, hBN have attracted much attention for its capability to exhibit various bright single-photon emitters at room temperature (RT) [16-18]. In addition, recent discoveries of optically addressable spin defects have further boosted the efforts to investigate these defects for quantum sensing applications [19-24]. In particular, the negatively charged boron vacancy ($V_B^-$) defect is of great interest due to its known molecular structure and reliable engineering methods [20,25-30].

The $V_B^-$ center consists of a boron vacancy surrounded by three nitrogen atoms and an extra electron captured from the environment (Fig. 1a). The spin-spin interaction along the out-of-plane direction splits the triplet ground state into $m_s$ = 0 and $m_s$ = ±1 manifolds with a zero-field splitting of $D_{GS}$ = 3460 MHz at RT (Fig. 1b) [20,31,32]. The ground state ODMR (GS-ODMR) contrast of ensemble $V_B^-$ centers can reach up to 46%, making it appealing for quantum applications [25]. The $V_B^-$ centers have been employed to sense temperature, pressure, and magnetic fields [33]. Thanks to the 2D nature of the host, these sensors are also used for the in-site imaging of the magnetic properties of layered materials by constructing van der Waals heterostructures [34]. The excited state of the $V_B^-$ centers is crucial in mediating the interaction between the electron spins and the nearby nuclear spins [35,36]. Although the excited state of this system has been predicted to be a triplet [31,32], the exact configurations of these energy levels are still unknown.

In this study, we investigated the excited state ODMR (ES-ODMR) spectrum of $V_B^-$ centers at cryogenic temperatures. The magnetic field-dependent ES-ODMR revealed an excited triplet state with a longitudinal splitting of $D_{ES}$ ~ 2160 MHz, and a g-factor of ~2. The $D_{ES}$ was also substantiated by approaching the ES $m_s$ = 0 and $m_s$ = -1 state under an external magnet, giving an emission minimum near the excited-state level anti-crossing (ESLAC) point at ~800 G (~2240 MHz).

For sample preparation, a straight gold stripline of 50 μm wide was first deposited on the Si/SiO$_2$ substrate before the exfoliated hBN was transferred to the stripline. This wide stripline ensures the generation of a homogeneous in-plane magnetic field at the center of the line for spin manipulations [25]. We then bombarded the hBN with Ga$^+$ ions at the center and edge of the gold line. The hBN flake on the stripline is shown in Fig. 1c (inset). The detailed sample preparation method is described in Supplementary Section A. Typical luminescence spectra of the $V_B^-$ ensembles at 293 K and 7 K are shown in Fig. 1c. The emission was collected into a multimode fiber via a home-built confocal microscope before being directed into a spectrometer or photon counting device. The stripline was wire-bonded to a chip carrier for microwave (MW) feeding through.

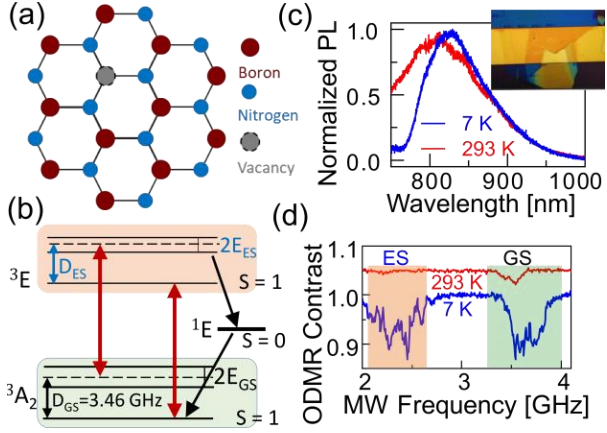

FIG. 1 (a) Schematic of a $V_B^-$ defect in hBN. (b) Energy structures of $V_B^-$ centers. It consists of a triplet ground state ($^3A_2$), a triplet excited state ($^3E$), and a singlet state ($^1E$). (c) Photoluminescence (PL) spectra of $V_B^-$ centers excited with 532 nm light at 293 K (red) and 7 K (blue). Inset: the microscopic image of the exfoliated hBN on a 50 μm wide gold stripline. (d) GS-ODMR (light green shadow) and ES-ODMR (light gold shadow) of the $V_B^-$ defects at 293 K and 7 K. The curves are vertically shifted for clarity.

The generation of $V_B^-$ centers was further confirmed with the GS-ODMR. The ODMR spectra were acquired by repeating the measurement cycles 100,000 times while recording the photon counts when the MW is periodically switched on ($I_{on}$) and off ($I_{off}$). Both the on and off durations last 10 μs each time (Fig. S1a). The contrast was calculated by $I_{on}/I_{off}$. The RT ODMR spectrum gives $D_{GS}$ = 3460 MHz and $E_{GS}$ = 49 MHz, as shown in Fig. 1d. Both the emission and ODMR spectra are in good agreement with other reports [20,26,37]. Compared to the RT ODMR spectrum, the center frequency of GS-ODMR ($D_{GS}$) blue-shifted to 3684 MHz at cryogenic temperatures, consistent with previous work [26,33]. Most importantly, at 7 K, we observed dips other than GS-splitting ranging from 2000 to 2600 MHz (light gold shadow in Fig. 1d), which we attribute to the zero-field splitting (ZFS) of the $V_B^-$ centers in the ES, as further confirmed by the following experiments.

We first confirm the nature of the ODMR dip around 2351 MHz by conducting pulsed ODMR measurement at 7 K (Fig. 2a). The critical point is to inject the MW after all populations relax into the ground state [35,38], and then compare the pulsed ODMR with the CW-ODMR. Experimentally, we first initialized most $V_B^-$ centers to the $m_s = 0$ state with a 5 μs laser pulse. We then apply a 1 μs MW pulse after a dwell time of 500 ns to ensure the complete relaxation of electrons to the GS (Fig. S1b). The spin state after MW operation is readout with another 5 μs laser pulse. Therefore, in pulsed ODMR, only the MW fields that are resonant with the GS-splitting can interact with the system and lead to an ODMR dip. Both CW-ODMR and pulsed ODMR resulted in a dip at the GS-splitting, while only CW-ODRM showed the dip at the ES-splitting (Fig. 2a), underscoring the nature of this dip around 2351 MHz as the ES-splitting.

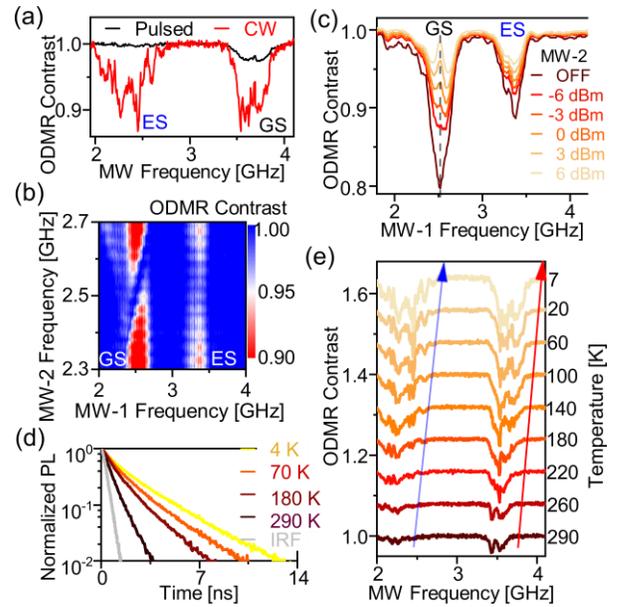

FIG.2 (a) ODMR spectra of $V_B^-$ centers under pulsed laser/MW excitation (black) and CW laser/MW excitation (red). (b) GS-ODMR and ES-ODMR spectra (under an external B-field of 400 G) while sweeping an additional MW-2. The discontinuous signals of GS- and ES-ODMR are due to the large step size of MW2. (c) ES-ODMR at different MW-2 power. The MW-2 frequency is fixed at 2520 MHz. (c) Lifetimes of $V_B^-$ centers at different temperatures. The IRF refers to the instrument response function that is much shorter than the lifetime. (d) Temperature-dependent GS- and ES-ODMR of $V_B^-$ centers. The two arrows guide the shifting splitting energy as varying the temperatures.

To confirm that the GS-ODMR and ES-ODMR signals are originated from the same defects, we performed a two MW experiment. Compared to the CW-ODMR, an additional MW2 is applied continuously to create spin mixing in the GS (Fig. S1c). This spin mixing would affect the ES-ODMR contrast. Since the ES and GS-ODMR fringes in Figure 2a are very broad, we applied a 400 G magnetic field to split the transitions between $m_s = 0$ and $m_s = \pm 1$ states, resulting in a GS transition ($0 \leftrightarrow -1$) near 2500 MHz and an ES transition ($0 \leftrightarrow +1$) near 3300 MHz. When the MW-2 is parked near resonance with the GS transition, the hole burning effect is observed for the GS-ODMR fringes; meanwhile, the ES-ODMR contrast is reduced due to the spin-depolarization of $m_s = 0$ (Fig. 2b). We also monitor how the ES-ODMR contrast changes while the MW-2 power is increased (Fig. 2c). At the highest power of MW-2, the ES-ODMR contrast is maximally reduced. These results indicate that the GS and ES signals in Fig. 2a are associated with the same type of defects.

The fluorescence lifetime of the ES is vital for ES-ODMR as a longer lifetime provides a longer time window for spin rotation in the ES [38]. Below we explain how the lifetime determines the ES-ODMR contrast with the help of the sequences for the ES-ODMR experiment. First, the spins of $V_B^-$ centers are mostly polarized by off-resonance pumping to the $m_s = 0$ GS. After spin-conserved off-resonance excitation, the MW starts to swap population between the $m_s = 0$ and $m_s = \pm 1$ states in the ES. The more population is transferred from the $m_s = 0$ state to the $m_s = \pm 1$ states, the more the PL intensity is reduced. Therefore, under weak MW power, a longer lifetime of ES can lead to more populations in the dark state, which will eventually lead to a more profound contrast in ES-ODMR.

To measure the relationship between the ES lifetime and the ES-ODMR contrast, we investigated them at different temperatures. Fig. 2d shows the ES lifetimes of the $V_B^-$ centers at different temperatures. The lifetime of ES can be extended from the RT value of 0.67 ns to 2.32 ns at 4 K. The enhancement factor is consistent with other reports [32]. This prolongment is likely due to the modification of nonradiative transitions.

We then measured the ODMR contrast as a function of temperature. The lowest GS-ODMR dip amplitude increases from 4% to 12 % when the temperature decreases from 297 K to 7 K (Fig. 2e). The $D_{GS}$ shows a blueshift of about 200 MHz while decreasing temperature. The temperature-dependent GS-ODMR contrast and $D_{GS}$ shift agree well with other reports [26]. Similar trends are observed for ES-ODMR, i.e., an increase of the lowest dip from ~2% at 290 K to 12% at 7 K. Moreover, the normalized areas formed by both GS and ES ODMR curves grow steadily during the cooling process (Table S1). We noticed that the ODMR contrasts in these measurements are limited by the available MW power rather than by physical limits. Unlike NV$^-$ centers in diamond, whose ES-ODMR quenches at 6 K due to the lack of dynamic Jahn-Teller effect [39], the $V_B^-$ centers exhibit the maximum ODMR contrast at the lowest achievable temperature. The GS-ODMR spectra exhibit fine structures due to the coupling between $V_B^-$ centers and nearby nitrogen nuclear spins [20]. Interestingly, the ES-ODMR also exhibits modulated fringes throughout the test temperatures. These fine structures may be related to hyperfine interactions between the electron and the nearby nuclear spins.

Next, we studied the response of $V_B^-$ centers to the external magnetic field at cryotemperature. Here, the external magnetic field is applied nearly perpendicular to the surface of the hBN. As shown in Fig. 3a, both GS-ODMR and ES-ODMR signals show the same slope as the B-field, inferring the same g-factor for the ES and GS. It indicates that the electron spin dominantly contributes to the g-factor while the orbital part contribution is negligible. The $D_{ES}$ is estimated by drawing a line cut at 400 G (Fig. 3b) and is attributed to the center frequency of two ES transitions (ES $m_s = 0$ to $m_s = \pm 1$ in Fig. 3b) at 2160 MHz. By increasing the magnetic field along the c-axis of the $V_B^-$ centers, the $m_s = -1$ state is brought close to the $m_s = 0$ state, resulting in turning points in both ES and GS [40,41]. Based on ODMR spectra, the turning points for the ES and GS are estimated to be 800 G and 1330 G (Fig. 3a), respectively, which agrees with the experimental results from $D_{ES}$ and $D_{GS}$.

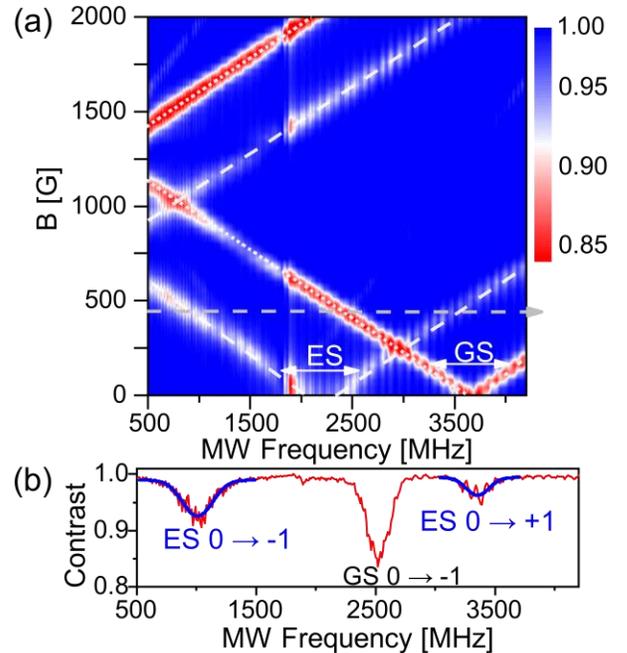

FIG.3 (a) Low-temperature B field-dependent GS- and ES-ODMR spectra. Dashed lines are the guidelines for ES-ODMR resonance frequency. The dotted line is the guide for GS. (b) Linecut ODMR spectra at 400 G, grey lines in (a). The blue curves are the Gaussian fit of the two ES transitions ($m_s = 0$ to $m_s = \pm 1$).

With the known parameters for g-factor and ZFS in the GS and ES, we then determined the eigenstates of $V_B^-$ centers by employing the Hamiltonian described in Supplementary Section E, neglecting interactions with nuclear spins [23]. When the external magnetic along the c-axis of the $V_B^-$ centers increases, the eigenenergy of $|-1\rangle$ states are getting close to the eigenenergy of |0⟩ (Fig. 4a shadow regions). This results in LAC points in the ES and GS (crossing points of black and red dotted lines in Fig. 4a). According to the calculation, the LAC in ES occurs at 769 G and 1307 G in GS, respectively (Fig. 4a).

To characterize the LACs, we conducted a MW-free experiment, by recording the emission intensity at different magnetic fields. Spin mixing between the bright ($m_s = 0$) and dark ($m_s = -1$) states in either ES or GS due to the

ESLAC at cryo-temperature [40]. The retained emission reduction of $V_B^-$ centers near the ESLAC at cryo-temperature is consistent with the observation of ES-ODMR. Both results indicate that the triplet ES of $V_B^-$ centers is maintained throughout the working temperature. This fact highlights the possibility of realizing DNP with $V_B^-$ defects even at cryotemperature.

We also looked into how the PL intensity changes at different misalignment angles. A larger misalignment angle

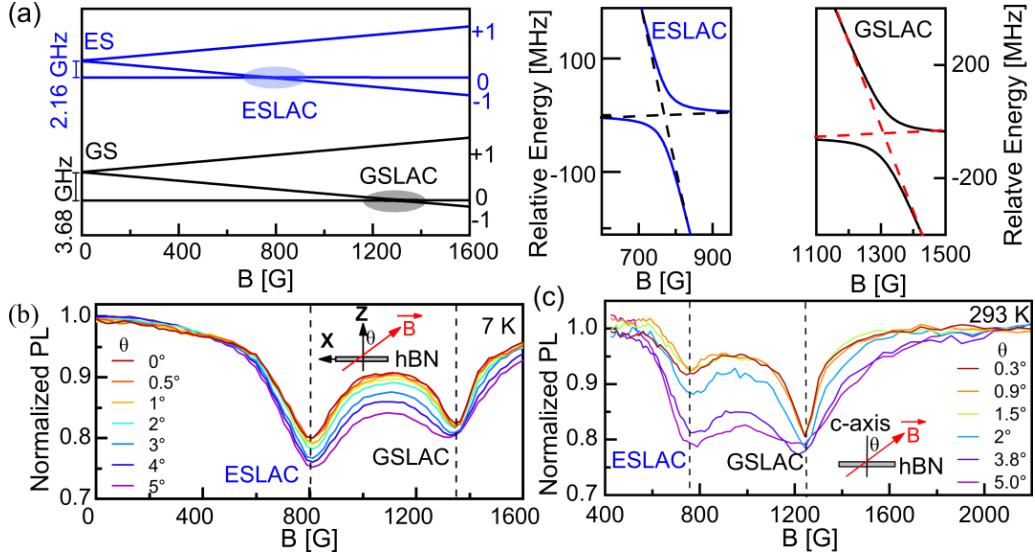

FIG. 4 (a) Energy level diagram of $V_B^-$ centers under a tilted external magnetic field (2 degrees) with respect to the c-axis of the sample. Two LAC magnetic fields between $m_s = 0$ and $m_s = -1$ are highlighted in the blue shaded ellipse for the ES and the black shaded ellipse for the GS. The hyperfine interactions are not included in the simulation. Inserts show the zoomed crossing points near ESLAC (middle) and GSLAC (right). The relative frequencies are obtained by deducting the frequencies at 769 G and at 1307 G. In the simulation, $D_{GS} = 3684$ MHz, $E_{GS} = 50$ MHz, and $D_{ES} = 2160$ MHz, $E_{ES} = 78$ MHz are employed. Normalized magnetic field-dependent emission count rate at 7 K (b) and 293 K (c). The emission count rate is normalized to emission at 0 G for (b) and at 2200 G for (c). In (b), the angle is defined between the magnetic field in the X-Z plane and Z-axis. There is a ***By*** component due to the 2 degree sample tilting in the Y-Z plane. In (c), the alignment uncertainty is 0.05 degree.

transverse magnetic field, strain, or hyperfine interaction [42-44], would reduce the emission rates since the spin populations are transferred from the bright to the dark state. The emission rates change is most evident when the energy difference between these two states is smallest, i.e., when the external magnetic fields are set to be near the ESLAC or GSLAC.

We investigated the magnetic field-dependent emissions at both 7 K and 293 K. At both temperatures, PL intensity drops are observed in the vicinity of the ESLAC and GSLAC (Fig. 4b and 4c). This temperature-dependent feature near ESLAC is in stark contrast to NV$^-$ centers in diamond. For NV$^-$ centers, the emission reduction near the ESLAC observed at room temperature disappears below 25 K due to a non-negligible contribution of the spin-orbit interaction [42,45], which complicated the DNP via

(a larger transverse magnetic field) generally leads to a more significant PL intensity reduction near the GSLAC and ESALC along with wider dips, which are also corroborated in Fig. 4b and 4c. However, all curves under different angles θ exhibited two local maximum PL reductions near the same magnetic fields, enabling the estimation of LAC magnetic fields in the ES and GS. We thus estimated $D_{ES} \sim 2240$ MHz and $D_{GS} \sim 3774$ MHz at 7 K, $D_{ES} \sim 2117$ MHz and $D_{GS} \sim 3472$ MHz at 293 K. Moreover, the magnetic field dependent emission spectra at 7 K exhibit less sensitivity to the tilting angle than those at 293 K. The magnitudes of the emission drops and the width of dips near the ESLAC and GSLAC show slight changes while tilting the angle varies from 0 to 5 degrees at 7 K. However, at 293 K, while the PL intensity drops near the GSLAC are almost unchanged, the corresponding widths changed drastically; by contrast, the drastic change near the ESLAC drops is the emission counts.

In conclusion, we have investigated both room- and low-temperature ES spectroscopies of the $V_B^-$ centers in hBN. The ES splitting $D_{ES}$ is estimated to be ~2160 MHz at 7 K and ~2117 MHz at 293 K. The ES-ODMR contrast is greater at cryogenic temperature due to longer fluorescence lifetimes at low temperatures. The ES is an important resource for manipulating nearby nuclear spins [36,46]. Knowledge of the energy levels of the ES provides the opportunity to realize DNP. The existence of ESLAC at cryotemperature provides a route towards nuclear spin manipulation at low temperatures.

Even though we know the ZFS of the ES, the realization of nuclear spin polarization via ESLAC remains a challenge. The potential challenges involve the smaller ratio of intersystem crossing rate from the ES manifolds to the singlet state [47,48], the unknown hyperfine interaction in the ES, the exact energy levels in the ES and the unknown spin-polarization mechanism. The symmetry of the $V_B^-$ centers system would be reduced due to the presence of the strain [31]. The symmetry dictates the energy structures of the $V_B^-$ centers and affects the spin-polarization mechanism [31]. Depending on whether electron spins are polarized into the $m_s = 0$ or $m_s = \pm 1$ states, the nuclear spin would be polarized into different states and the degree of polarization would be affected. In this regard, resonant optical addressing of a single $V_B^-$ center could be vital to reveal the detailed energy levels in the ES and to clarify the interaction between nuclear spins and electron spin [39]. Moreover, the coherent manipulation of the ES would shed light on the spin-dependent relaxation rates from the ES [49]. After knowing these, the nuclear spin polarization could be conceived.

**Acknowledgments**

We acknowledge Singapore National Research foundation through QEP grant (NRF2021-QEP2-01-P02, NRF2021-QEP2-03-P01, 2019-0643 (QEP-P2) and 2019-1321 (QEP-P3)) and Singapore Ministry of Education (MOE2016-T3-1-006 (S)), the Australian Research council (via CE200100010), the Asian Office of Aerospace Research and Development grant FA2386-17-1-4064.